\documentclass[prb, aps, amssymb, twocolumn, floatfix
 ]{revtex4}
\usepackage{graphicx}
\begin{document}
\title{Temperature measurement at the end of a cantilever using oxygen
paramagnetism in solid air}
\author{Kent R. Thurber}
\affiliation{U.S. Army Research Laboratory, AMSRL-SE-EM, Adelphi,
Maryland 20783}
\author{Lee E. Harrell}
\affiliation{Department of Physics, U.S. Military Academy, West
Point, New York 10996}
\author{Doran D. Smith}
\affiliation{U.S. Army Research Laboratory, AMSRL-SE-EM, Adelphi,
Maryland 20783}
\date{\today}
\begin{abstract}
We demonstrate temperature measurement of a sample attached to the
end of a cantilever using cantilever magnetometry of solid air
``contamination'' of the sample surface.  In experiments like our
Magnetic Resonance Force Microscopy (MRFM), the sample is mounted
at the end of a thin cantilever with small thermal conductance.
Thus, the sample can be at a significantly different temperature
than the bulk of the instrument.  Using cantilever magnetometry of
the oxygen paramagnetism in solid air provides the temperature of
the sample, without any modifications to our MRFM (Magnetic
Resonance Force Microscopy) apparatus.
\end{abstract}
\maketitle

In many Magnetic Resonance Force Microscopy
(MRFM)\cite{GaAsMRFM,MRFM1,CaF,MRFM2,MRFM3,MRFM4} and cantilever
magnetometry experiments,\cite{magnetometry} the sample being
measured is mounted at the end of a thin cantilever. The
cantilever often has very poor thermal conductance because of its
small cross section in comparison to its length.  Even a small
heat input into the sample can raise its temperature significantly
from the temperature at the opposite end (base) of the cantilever.
Obviously, the temperature of the sample can be critical to the
results of an experiment because of many temperature dependent
factors like spin polarization or time constants.\cite{MRFM1,CaF}
Mounting a temperature sensor at the end of the cantilever would
be difficult, requiring a sensor of $\leq$10 $\mu$m dimensions
integrated with the cantilever.  An indirect way to measure the
temperature of a cantilever is to measure the thermal (Brownian)
motion of the cantilever.\cite{brownian}  Using simple Brownian
motion requires that the spring constant of the loaded cantilever
is known and that the temperature is uniform along the cantilever
and equal to the sample temperature.  In addition, the Brownian
motion method relies on the absence of non-thermal sources of
cantilever motion for correct temperature measurement. Sources of
sample heating (such as RF magnetic fields and optical power) can
also be sources of non-thermal motion.

A simple alternative is to use cantilever magnetometry to measure
the paramagnetism of oxygen in frozen air on the sample.  This
technique does not require any modification of the experimental
probe used for MRFM or cantilever magnetometry, and provides a
relative measure of the sample temperature up to the sublimation
temperature of the solid air.  In essence, the temperature is
measured by using the temperature dependence of the magnetic
susceptibility of the air ``contamination" of the sample surface.
The paramagnetic electron spins of the oxygen molecules in the
solid air have a large, easily measureable temperature dependence
of magnetic susceptibility.  This technique could also be done
just using the susceptibility of the sample itself if there is
enough temperature dependence of the susceptibility.


The experimental setup is the same as used for our sample on
cantilever MRFM experiments.\cite{GaAsMRFM}  The sample for MRFM
(in this case GaAs) is mounted with thermally conductive silver
epoxy on a gold-coated SiN$_{x}$ cantilever with spring constant,
k$\sim$0.05 N/m.   The SiN$_{x}$ cantilever has very poor thermal
conductance because of the low thermal conductivity of SiN$_{x}$
and thin cross section (the cantilever is 0.6 $\mu$m thick, 22
$\mu$m wide, and 320 $\mu$m long).\cite{cantilever} In order to
increase the thermal conductivity of the cantilever, it was coated
on each side with either 300 \AA/700 \AA\ or 300 \AA/1700 \AA\ of
Ti/Au. Coating on both sides is required to reduce bending of the
cantilever from the differential contraction of the gold and
SiN$_{x}$ on cooling. The position of the cantilever is measured
with an optical interferometer.\cite{CaF}  Below the sample is the
source of the magnetic field gradient, a 250 $\mu$m diameter iron
cylinder. The magnetic field gradient ($\delta B_{z}/\delta z =
6000$ to 9000 T/m) applies a static force,
\begin{equation}
F_{z} = \mu_{z} \frac{\delta B_{z}}{\delta z}
\end{equation}
to the net magnetic moment of the oxygen electrons, $\mu$, which
we measure by the deflection of the cantilever.  Raw data from the
interferometer is shown in figure 1. As the magnetic field is
increased, the oxygen electrons become more polarized and the
force between the oxygen and the magnetic field gradient becomes
larger.  The cantilever deflects towards the iron magnet. Raising
the temperature decreases the susceptibility and thus the amount
of deflection (figure 1(c)).

\begin{figure}
\includegraphics{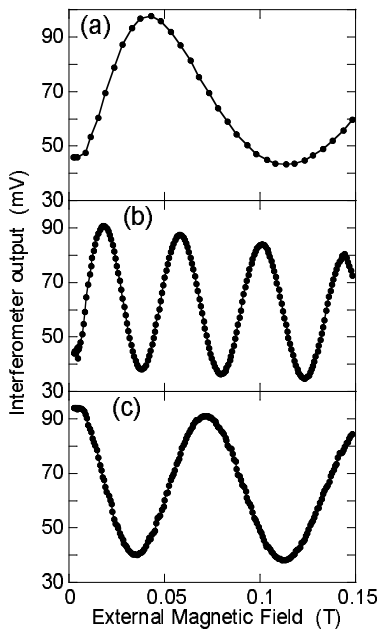} \caption{Optical interferometer measurement
of the cantilever position as a function of external magnetic
field for (a) 4.2 K without air frozen on sample, (b) 4.2 K with
frozen air on sample, and (c) 8.4 K with frozen air on sample.
Each interferometer oscillation represents $\lambda/2=390$ nm of
cantilever deflection.} \label{fringes}
\end{figure}

This technique of temperature measurement could be done with the
sample itself, if the sample has sufficient temperature dependence
of its magnetic susceptibility for easy measurement. However,
``contamination" of the sample and cantilever surface with solid
air can be used with an arbitrary sample.  In addition, air is a
resource available in almost every lab.  Coating the sample with a
small amount of frozen air can easily be done by not fully pumping
out the probe before cool down. Because of the sensitivity of the
cantilever, very little air is required. Just 10$^{11}$ oxygen
molecules ($\sim$monolayer) will result in several interferometer
fringes of deflection (hundreds of nm) during a magnetic field
sweep to 9 T. Deliberately allowing air into the vacuum system
with a non-regulating valve typically resulted in too much air
contamination and the magnetic force would stick the cantilever to
the iron magnet.

This provides a cautionary warning for cantilever magnetometry or
MRFM experiments.  Oxygen contamination must be avoided for
sensitive magnetometry or MRFM experiments because of the large
magnetic susceptibility of oxygen.  For this reason, the
temperature measurements were done first and then the air
contamination removed before the MRFM experiments. The air
contamination could be removed by heating the probe head to about
20 K while pumping. The best technique for avoiding air
contamination was to use the heater to keep the probe head warm,
while the rest of the probe was cooled down.

We should also note that solid air is a better choice for large
and simple temperature dependence of the magnetic susceptibility
than pure oxygen. Pure solid oxygen has antiferromagnetic
interactions between the molecules and undergoes two phase
transitions in the solid state at low temperatures (43.8 and 23.9
K).\cite{EPR,adsorbed}  By contrast, the oxygen electrons in solid
air have been studied recently with EPR at 5 K and shown to behave
as $S=1$ paramagnets\cite{EPR} following the Hamiltonian
\begin{equation}
H = g \mu_{B} B S_{z} + D \left( S_{z}^{2} - \frac{S(S+1)}{3}
\right)
\end{equation}
where $g=2.0$ and the zero field splitting $D=5.1$ K.  The
Hamiltonian is written for the case where the axis of the oxygen
molecule is along the magnetic field.  In the frozen air, the
oxygen molecules have a distribution of their molecular axes with
respect to the magnetic field.\cite{EPR}  Apparently, the oxygen
molecules are dispersed enough among the nitrogen molecules in the
solid air to prevent antiferromagnetic ordering. Similarly, dilute
adsorbed oxygen results in roughly Curie-Weiss law behavior of the
susceptibility, $1/(T-\theta)$, with $\theta$ small and
negative.\cite{adsorbed} Figure 2 shows the deflection of the
cantilever up to high field. The saturation of the susceptibility
at high field is consistent with spin 1 and $g\approx2$ electronic
moments.  To provide a calibration for this temperature
measurement method, we heated the probe head, which results in a
nearly uniform temperature of the probe head, cantilever, and
sample.  The probe head itself has its temperature measured with a
resistive sensor.\cite{tsensor}  As shown in figure 3, the
temperature dependence of the magnetic susceptibility does indeed
roughly follow a Curie-Weiss law with the fitted $\theta=-2$ K.

\begin{figure}
\includegraphics{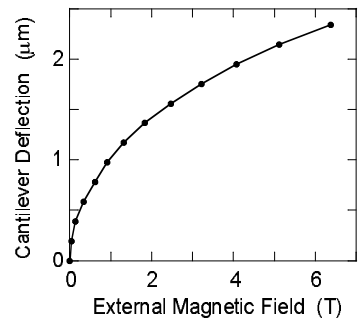} \caption{Cantilever deflection as a function
of external magnetic field for air ``contaminated'' sample at 4.2
K.} \label{displacement}
\end{figure}

\begin{figure}
\includegraphics{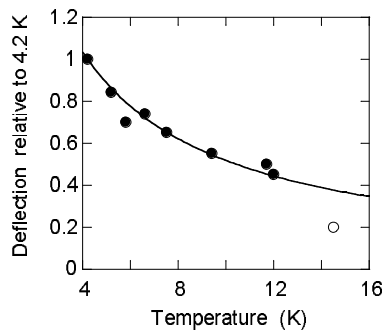} \caption{Oxygen magnetometry as a function
of temperature. The oxygen magnetometry, $\bullet$, is plotted as
the deflection of the cantilever over a magnetic field sweep
(0.15, 0.8, or 2 T) relative to the deflection over the field
sweep at 4.2 K. At the highest temperature ($\circ$ 14.5 K),
oxygen desorbs when under low pressure and the magnetometry is no
longer reversible with temperature. The line is a Curie-Weiss law
fit to the data below 14.5 K ($\theta=-2$ K)} \label{curie}
\end{figure}

Then, we applied this technique to measure the sample temperature
under conditions when it is different from the temperature of the
rest of the probe head.  The magnetic susceptibility of the sample
with air ``contamination'' is measured by the deflection of the
cantilever by a change in magnetic field.  The temperature is
derived by comparison with the calibration done by heating the
entire probe head.  The calibration step could be avoided by
assuming the Curie-Weiss law temperature dependence of the oxygen
susceptibility and the dominance of the oxygen susceptibility over
any other magnetic susceptibility of the sample or cantilever.  In
these MRFM experiments,\cite{GaAsMRFM} the GaAs sample can be
heated from two major sources. First, we are applying RF magnetic
field which can heat the sample and gold-coated cantilever
directly by eddy current heating and indirectly through gas
conduction. Second, and more significantly, we are shining light
on the sample for optical pumping.\cite{OP} Another smaller heat
source is the optical power used in the interferometer
measurement.  We measured the susceptibility and found, for
instance, that roughly 4 Gauss of RF (51.5 MHz) magnetic field
raised the sample temperature about 1 K. Even more dramatic is the
effect of optical pumping (Fig. 4).  The $\sim$80 $\mu$W of
optical power on the GaAs sample at the end of the cantilever can
heat it by 3 K, while the main bulk of the probe has warmed by
less than 0.1 K. This local heating is consistent with the
estimated thermal conductance of the cantilever, $\approx$25
$\mu$W/K, which is dominated by the 700 \AA\ gold coating.

\begin{figure}
\includegraphics{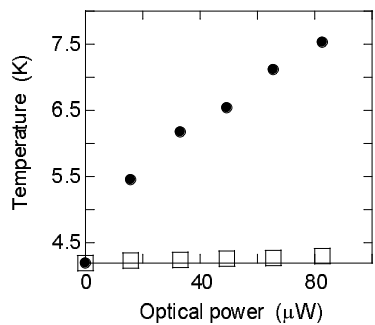} \caption{Comparison of sample temperature
measured by oxygen magnetometry ($\bullet$) and the temperature of
the bulk of the probe ($\square$) while shining 823 nm light on
the sample. Cantilever is coated on both sides with 300 \AA\ Ti
and 700 \AA\ Au.} \label{optical}
\end{figure}

We have presented a technique for measuring the temperature of a
sample at the end of a cantilever.  In general, oxygen
contamination must be avoided for cantilever magnetometry and MRFM
experiments.  But here, by measuring the susceptibility of the
oxygen in solid air contamination, we can determine the
temperature without any contact to the sample.  This technique can
be extended by using the susceptibility of some other
``contaminant" or just the sample itself.  In addition, these
measurements are done without any modification of the apparatus
which we use primarily for MRFM measurement of the nuclear
polarization of the sample.

This work was supported by the DARPA Defense Science Office Spins
in Semiconductors program. The cantilevers were coated by Monica
Taysing-Lara. The authors would like to thank John A. Marohn, John
Sidles, and Dan Rugar for many helpful discussions.


%
%

%
%

\end{document}